\documentclass[a4paper,fleqn]{cas-dc}
\usepackage[utf8]{inputenc}
\usepackage{textgreek}
\usepackage[utf8]{inputenc}
\usepackage[T1]{fontenc}
\usepackage{newunicodechar}
\newunicodechar{ȩ}{\k{e}}

\usepackage[authoryear,longnamesfirst]{natbib}
\usepackage{float}
\usepackage{caption}
\usepackage{subcaption}
\usepackage[dvipsnames]{xcolor}
\usepackage{hyperref}
\usepackage{tabularx}
\usepackage{booktabs}
\usepackage{cuted}
\usepackage{array}
\usepackage{algorithm}
\usepackage{algpseudocode}

\def\tsc#1{\csdef{#1}{\textsc{\lowercase{#1}}\xspace}}
\tsc{WGM}
\tsc{QE}

\begin{document}
\let\WriteBookmarks\relax
\def\floatpagepagefraction{1}
\def\textpagefraction{.001}

\shorttitle{}    

\shortauthors{}  

\title [mode = title]{Targeting Behavior or Intention? Intervention Strategies in Expressed--Private Opinion Dynamics}  



%














\author[1]{Barbara Kami\'{n}ska}[orcid=0000-0003-0663-8753]


\ead[1]{b.kaminska@pwr.edu.pl}
\credit{Conceptualization, Methodology, Formal analysis, Investigation, Software, Visualization, Writing}

\author[1]{Arkadiusz J\k{e}drzejewski}[orcid=0000-0002-7965-2014]


\ead[2]{arkadiusz.jedrzejewski@pwr.edu.pl}
\credit{Methodology, Validation, Investigation, Writing - original draft, Writing - review \& editing}%

\author[1,2]{Katarzyna Sznajd-Weron}[orcid=0000-0002-1851-8508]
\cormark[3]
\ead[3]{katarzyna.weron@pwr.edu.pl}

\credit{Conceptualization, Funding acquisition, Project administration, Supervision, Writing – original draft, Writing – review \& editing.}
\cortext[cor3]{Principal corresponding author}

\affiliation[1]{organization={Wroclaw University of Science and Technology, Faculty of Management, Department of Computational Social Science},
    addressline={Wyb. Wyspia\'{n}skiego 27},
    city={Wroclaw},
    postcode={50-370},
    country={Poland}}

\affiliation[2]{organization={Complexity Science Hub Vienna},
    city={Vienna},
    country={Austria}}



\begin{abstract}
Exogenous interventions, including economic incentives, can be used to promote behavioral change and the adoption of new products or practices. They may act on different levels, either directly affecting customers’ behavior (expressed level) or changing intentions (private level). The key question is which intervention strategy is more effective: targeting the expressed or the private level. Although expressed-private opinion (EPO) models are especially suitable for addressing this problem, they have not yet been used in this context. To fill this gap, we extend previous agent-based EPO models by incorporating exogenous interventions and compare different intervention strategies. We analyze the resulting model using a mean-field approximation, complemented by Monte Carlo simulations on a complete graph. Our results show that there is no universal answer to whether an intervention should target the expressed or the private opinion. The effectiveness of an intervention depends on the adoption level at the moment of its introduction. Interventions acting on the expressed level are more effective when adoption is still low, whereas interventions acting on the private level are more effective when adoption is already high and the goal is to support internal acceptance and reduce dissonance. Moreover, the model shows that high adoption is accompanied by low dissonance between expressed and private opinions, which is consistent with self-congruity theory. This agreement provides theoretical support for the model and strengthens its policy implications.
\end{abstract}


\begin{highlights}
    \item Exogenous intervention is introduced into an Expressed-Private Opinion model.
    \item Interventions on private and expressed opinions are compared.
    \item The best intervention strategy depends on the current adoption level.
    \item Expressed-level intervention promotes adoption when adoption is low.
    \item High adoption with low dissonance supports the self-congruity interpretation.
\end{highlights}



\maketitle

\section{Introduction}
\label{sec:introduction}
Would a customer be more likely to bring a reusable shopping bag because single-use bags are no longer free, or because education have changed how they think about disposable plastics? The case of plastic bags illustrates a broader problem in the study of interventions: the same behavioral goal can be approached through different mechanisms. On the one hand, economic incentives can directly modify behavior by changing the costs associated with particular choices. A well-known example is the Irish plastic bag levy, which led to a dramatic reduction in plastic bag use in retail outlets, estimated at about 90\% \citep{convery_most_2007}. On the other hand, behavior-change interventions such as nudges, norm messages, and education offer ways to reduce demand for single-use plastics by influencing perceptions, norms, and individual evaluations \citep{truelove_curbing_2022}. 

The plastic-bag example is one of many cases showing that different types of interventions can change how people act. Similar effects have been observed in other environmental contexts: social influence can support resource conservation \citep{abrahamse_social_2013}, and information about environmental impacts can promote conservation behavior \citep{delmas_information_2013}. Beyond environmental behavior, subsidies can encourage healthier food choices \citep{an_effectiveness_2013}. These examples show that interventions may operate through different mechanisms: by changing the social context, altering economic incentives, or reshaping beliefs and evaluations. This leads to a central question: should an intervention target behavior itself, or internal acceptance?

To address this question, we use an agent-based model that belongs to the class of expressed-private opinion (EPO) models \citep{dong_opinion_2024,kaminska_impact_2025}. In this framework, each agent is described by two opinion variables: an expressed opinion, which is visible to others and can be interpreted as a public declaration or behavior, and a private opinion, which represents internal acceptance or intention. This distinction is consistent with empirical work on the attitude--intention--behavior gap in green consumption \citep{elhaffar_towards_2020}, the attitude--behavior gap in sustainable consumption \citep{park_exploring_2020}, and the intention--behavior gap in domains such as sustainable clothing \citep{rausch_bridge_2021}, sustainable tourism \citep{nieto-garcia_consumer_2024}, green food purchasing \citep{de_sio_product_2024}, and meat reduction \citep{linder_psychological_2026}. It is also central to the present problem because an intervention may change what agents do or publicly express without necessarily changing what they privately think; conversely, it may change private attitudes without immediately producing visible behavioral change. 

The paper is organized as follows. In Sec.~\ref{sec:model}, we define the EPO agent-based model with an exogenous intervention and specify the update rules for private and expressed opinions. In Sec.~\ref{sec:methods}, we describe the Monte Carlo simulations and the mean-field approximation used to analyze the model. In Sec.~\ref{sec:results}, we present the results, focusing in particular on stationary adoption levels and dissonance patterns obtained when interventions are applied to the private or expressed level. Then, in Sec.~\ref{sec:discission}, we discuss the results in relation to consumer behavior, self-congruity theory, and the intention--behavior gap. Finally, in Sec.~\ref{sec:conclusions}, we conclude with simple economic policy implications and discuss the limitations of our approach.

\section{The model}
\label{sec:model}
We consider a system of $N$ agents placed on the nodes of an arbitrary graph, where interactions are allowed only between directly connected agents. Throughout this paper, we use the pronoun \textit{it} when referring to an agent, rather than \textit{he} or \textit{she}, to emphasize that an agent is a virtual entity in the algorithm, not a human being. Analogously to \cite{kaminska_impact_2025}, the state of each agent $i=1,\ldots,N$ is described by a pair of binary, time-dependent variables, $S_i(t)=\pm 1$ and $\sigma_i(t)=\pm 1$. For brevity, we often write them as $S_i$ and $\sigma_i$, indicating the time dependence explicitly only where needed. The first variable, $S_i$, denotes the expressed, or public, opinion. It may also be interpreted as a behavior, that is, a state observed by other agents. The second variable, $\sigma_i$, represents the private opinion, known only to the given agent.

The binary form is natural for describing opinions such as \textit{yes} or \textit{no}, being \textit{for} or \textit{against} a given issue, or, in the context considered here, adopting or not adopting a certain product, practice, or behavior. We use the following notation: $\uparrow$ denotes a positive opinion, $+1$, whereas $\downarrow$ denotes a negative opinion, $-1$. Since the state of an agent is described by a pair of expressed and private opinions, the state of an agent with $S_i=1$ and $\sigma_i=-1$ can be written compactly as $\uparrow\downarrow$ ($S\sigma$ - first arrow stands for the expressed opinion, second for the private). Consequently, each agent can be in one of four possible states: $\uparrow\uparrow$, $\uparrow\downarrow$, $\downarrow\uparrow$, or $\downarrow\downarrow$.

\begin{figure*}[htbp]
    \centering
    \includegraphics[width=0.7\textwidth]{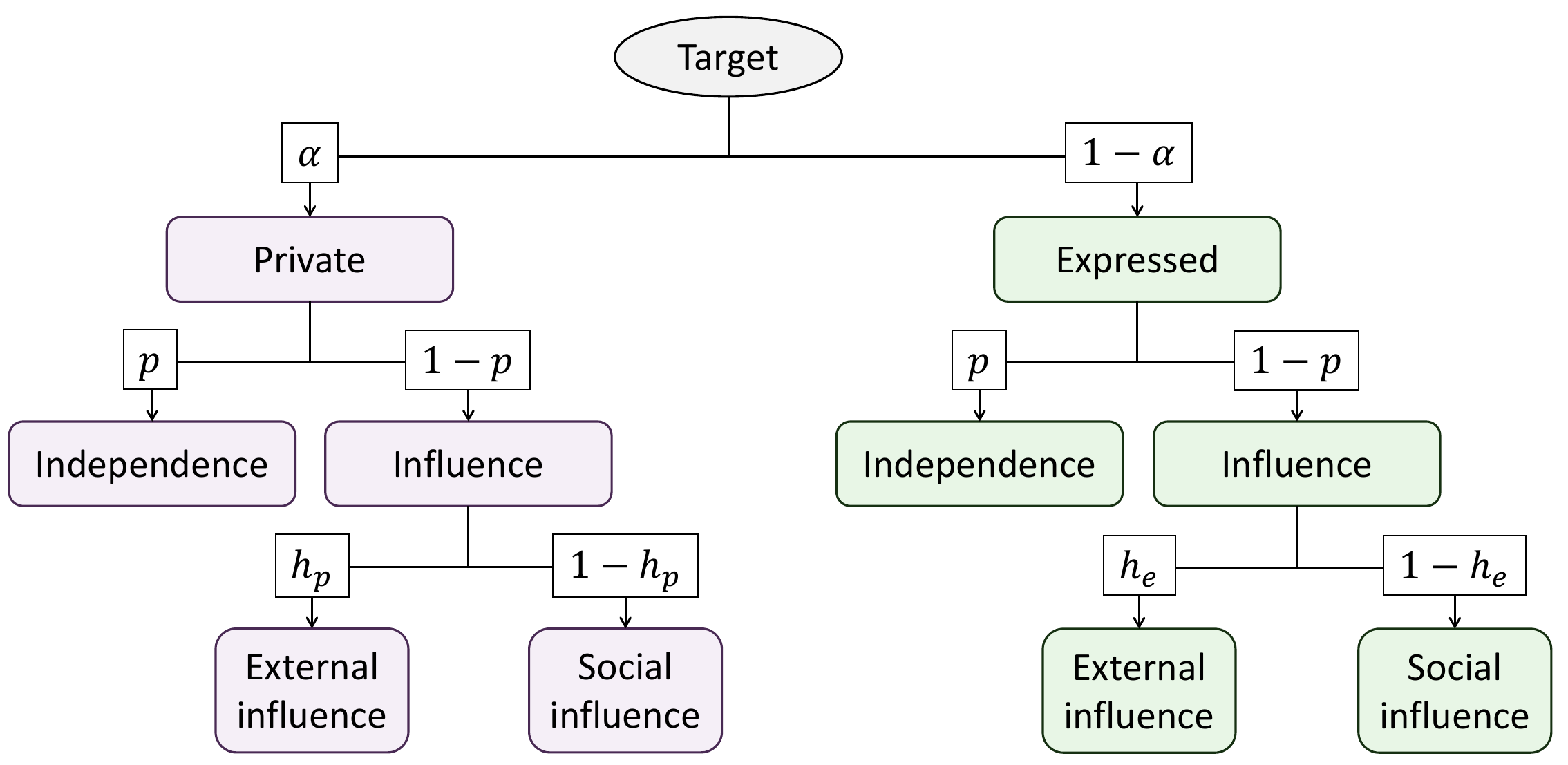}
    \caption{\textbf{Schematic representation of the model.} The first branching separates private and expressed opinion updates, which occur with probabilities $\alpha$ and $1-\alpha$, respectively. The target agent is then classified as independent or susceptible to influence. If influence occurs, its source is selected as either social or external. Setting $h_p=h_e=0$ reduces the model to the one considered in \cite{kaminska_competition_2026}.}
    \label{fig:EPO_field_a}
\end{figure*}

Agents' opinions change due to two basic mechanisms, independence and conformity, both motivated by the four-dimensional model of social response \citep{nail_proposal_2000}. In the absence of intervention, conformity is driven only by interactions with other agents \citep{jedrzejewski_think_2018,kaminska_impact_2025,kaminska_competition_2026}. In this work, we introduce an additional source of influence, namely an exogenous intervention. Thus, a target agent may conform either to other agents or to the external source, as shown in Fig.~\ref{fig:EPO_field_a}. Below, we specify how independence, peer influence, and exogenous intervention act on the private and expressed levels.

Independence captures situations in which an agent acts without taking into account what others do or say. On the expressed level, this means that the agent reveals its private opinion, that is, $S_i := \sigma_i$, where $:=$ denotes that the variable on the left-hand side is updated to the value on the right-hand side. On the private level, we associate independence with rethinking a given issue. Following earlier EPO models, which provide the starting point for the present work, we implement private independence as a random change to the opposite opinion; that is, with probability $1/2$, we set $\sigma_i := -\sigma_i$ \citep{jedrzejewski_think_2018,kaminska_impact_2025,kaminska_competition_2026}.

Conformity refers to situations in which an agent takes the opinion of the source of influence. In the present model, there are two possible sources of influence: a group of peers and an external intervention. We first describe peer influence. As in the broader class of $q$-voter-based models \citep{castellano_statistical_2009}, the peer source is a group of $q$ agents. We call this group the influence group, or $q$-panel. This assumption reflects the fact that although the total number of peers in an agent's social network may be much larger, only a limited number of them are involved in a given interaction, for example family members, friends, or workmates with whom the agent happens to discuss the issue at a given moment. The size of this group is fixed and equal to $q$, whereas its composition may change from one update to another.

On the expressed level, peer influence may lead to two types of conformity, depending on the initial state of the target agent. If the target is in harmony, $S_i=\sigma_i$, it is less susceptible to social influence. Therefore, only a unanimous $q$-panel can make the target change its expressed opinion. In this case, the target adopts the opinion expressed by the $q$-panel. This type of response is called compliance \citep{nail_proposal_2000}. If the target agent is initially in dissonance, $S_i\neq\sigma_i$, it is more willing to change its expressed opinion. Then, it is sufficient that at least one agent in the $q$-panel expresses the opinion already held privately by the target agent. In this case, the target reveals its private opinion. This type of response is called disinhibitory contagion \citep{nail_proposal_2000}.

We implement conformity-driven changes of the expressed opinion as follows. First, we randomly select $q$ neighbors of the target agent. Then:
\begin{itemize}
\item if $S_i \neq \sigma_i$, disinhibitory contagion applies: if there exists at least one neighbor $i_j$, $j=1,\ldots,q$, such that $S_{i_j}=\sigma_i$, then $S_i := \sigma_i$;
\item if $S_i = \sigma_i$, compliance applies: if all selected neighbors express the same opinion, $S_{i_1}=\ldots=S_{i_q}$, then $S_i := S_{i_1}$.
\end{itemize}

On the private level, peer influence can only reduce cognitive dissonance, in line with \cite{kaminska_impact_2025}. Thus, the private opinion of the target agent changes only if the selected $q$-panel is unanimous and expresses the same opinion as the target agent's expressed opinion. In other words, if $S_i = S_{i_1}= \ldots = S_{i_q}$, then $\sigma_i := S_i$.

The external intervention is implemented in a simpler way. In contrast to peer influence, it does not depend on the state of the $q$-panel. It represents an external source promoting adoption, that is, the positive state. Therefore, if the selected opinion level follows the intervention, the corresponding opinion is set to $+1$. On the expressed level, we set $S_i := 1$, whereas on the private level, we set $\sigma_i := 1$.

Finally, as in \cite{kaminska_competition_2026}, we assume that during a single elementary update an agent can change only one opinion level: the private opinion with probability $\alpha$ and the expressed opinion with probability $1-\alpha$. The rules for independence, peer influence, and external intervention were described above, while Fig.~\ref{fig:EPO_field_a} summarizes the sequence of choices in a single update. For an algorithmic representation of the model, see Algorithm~\ref{alg:opinion_update} in Appendix~\ref{app:algorithm}.

Summarizing, the model considered here combines elements of two previous EPO models. As in \cite{kaminska_impact_2025}, conformity on the private level is allowed only when it reduces cognitive dissonance. As in \cite{kaminska_competition_2026}, private and expressed opinions may evolve at different rates. In \cite{kaminska_competition_2026}, both versions of the model were considered, with and without cognitive-dissonance reduction. Here, we use the former version, which is more appropriate for the pro-environmental context considered in this paper. We also checked that removing this mechanism does not change the qualitative role of the intervention. Therefore, for clarity, we present only the results for the version with cognitive-dissonance reduction. The new element introduced in the present work is an external intervention, which can act either on the expressed or on the private level. The comparison of these two intervention strategies is the main focus of the paper.

\section{Methods}
\label{sec:methods}
We describe the macroscopic state of the system by the fractions of agents in the four possible states $X \in \{{\uparrow \uparrow, \uparrow \downarrow, \downarrow \uparrow, \downarrow \downarrow\}}$:
\begin{flalign}
    c_X(t)=\frac{N_X(t)}{N}, \label{eq:def_concentration}
\end{flalign}
where $N_X(t)$ denotes the number of agents in state $X$ at time $t$. Since each agent is always in exactly one of the four states, these fractions satisfy the normalization condition:
\begin{flalign}
    c_{\uparrow \uparrow}(t) + c_{\uparrow \downarrow}(t) + c_{\downarrow \uparrow}(t) + c_{\downarrow \downarrow}(t) = 1. \label{eq:normalization}
\end{flalign}

Because of the normalization condition \eqref{eq:normalization}, only three of the four fractions $c_X$ are independent. Following \cite{jedrzejewski_think_2018,kaminska_impact_2025,kaminska_competition_2026}, instead of using three selected fractions $c_X$, we characterize the macroscopic state of the system by three aggregate variables:
\begin{flalign}
c_S(t) = c_{\uparrow \uparrow}(t) + c_{\uparrow \downarrow}(t),
\label{eq:def_Ce}
\end{flalign}
which denotes the fraction of agents with a positive expressed opinion,
\begin{flalign}
c_{\sigma}(t) = c_{\uparrow \uparrow}(t) + c_{\downarrow \uparrow}(t),
\label{eq:def_Cp}
\end{flalign}
which denotes the fraction of agents with a positive private opinion, and
\begin{flalign}
d(t)= c_{\uparrow \downarrow}(t) + c_{\downarrow \uparrow}(t),
\label{eq:def_Cd}
\end{flalign}
which denotes the fraction of agents in dissonance.

To determine the macroscopic variables defined above, we use Monte Carlo (MC) simulations, described in Sec.~\ref{sec:MC}, and analytical calculations within the mean-field approximation (MFA), described in Sec.~\ref{sec:MFA}. Although MC simulations can, in principle, be used to study the model on arbitrary network structures, here we focus on the complete graph. In this topology, each agent can interact with any other agent with equal probability, which makes the MC results directly comparable with the MFA predictions. Moreover, the complete graph does not introduce additional parameters associated with network topology. This allows us to isolate the effect of the new ingredient considered here, namely the external intervention. Finally, using MC simulations and MFA in parallel provides an independent consistency check of the results.

\subsection{Monte Carlo simulations}
\label{sec:MC}
We perform MC simulations by iterating the elementary update described generally in Sec.~\ref{sec:model} and more precisely in Algorithm~\ref{alg:opinion_update} in Appendix~\ref{app:algorithm}. Each simulation starts from a given initial configuration. Unless stated otherwise, all agents are initially in harmony, that is, $S_i(0)=\sigma_i(0)$ for $i=1,\ldots,N$. We vary the initial fraction of positive opinions to check whether, after a transient period, the system reaches the same long-run (stationary) state or whether different initial conditions lead to different stationary states.

We measure time in Monte Carlo steps (MCS). One MCS consists of $N$ elementary updates, so that, on average, each agent has one opportunity to update its state during a single MCS. For each set of parameters and initial conditions, we run the simulation for $8000$ MCS. We discard the first $4000$ MCS as a transient period and use the remaining $4000$ MCS to estimate the stationary values of the macroscopic variables $c_S$, $c_\sigma$, and $d$ by time averaging.

We choose the length of the transient period on the basis of the observed time evolution of the system. As shown in Fig.~\ref{fig:trajectories}, the trajectories reach stationary regimes well before the averaging window begins. Since we study the model on a complete graph, we do not need to average over network realizations. For more complex network structures, such averaging would be necessary because the results could depend on the particular network topology.

All Julia codes used for the MC simulations are available on Zenodo \citep{kaminska_2026_zenodo}

\subsection{Mean-field approximation}
\label{sec:MFA}
We use random sequential updating, meaning that in a single elementary update, we update the state of one randomly chosen agent rather than synchronously updating all agents. Therefore, during one elementary update, which corresponds to $\Delta t = 1/N$, the fraction of agents in each of the four states may increase by $1/N$, decrease by $1/N$, or remain unchanged. Following \cite{jedrzejewski_think_2018,kaminska_impact_2025}, we denote the probabilities of these changes as:
\begin{flalign}
\gamma_X^+ & = \text{Prob}\left(c_X(t + \Delta t) =  c_X(t) + \frac{1}{N}\right), \nonumber \\
\gamma_X^- & = \text{Prob}\left(c_X(t + \Delta t) =  c_X(t) - \frac{1}{N}\right),  \nonumber \\
\gamma_X^0 & = \text{Prob}\left(c_X(t + \Delta t) =  c_X(t)\right) = 1 - \gamma_X^+ - \gamma_X^-.
\label{eg:gammas}
\end{flalign}
Using these probabilities, we write the discrete-time evolution of $c_X$ as:
\begin{flalign}
    c_X(t + \Delta t) &= c_X(t) +  \frac{\gamma_X^+ - \gamma_X^-}{N} \nonumber \\
    & = c_X (t) + \left(\gamma_X^+ - \gamma_X^-\right)\Delta t. 
    \label{eq:rate_discrete}
\end{flalign}

In the limit $N\rightarrow\infty$, we have $\Delta t\rightarrow 0$, and the discrete-time evolution becomes a system of differential equations:
\begin{flalign}
    \frac{dc_X}{dt} &= \gamma_X^+ - \gamma_X^-. \label{eq:rate}
\end{flalign}

Explicit formulas for the system of equations \eqref{eq:rate} are given in Appendix~\ref{app:MFA}.

In the stationary state, the fractions of agents in the four states are constant. Hence, they satisfy the condition:
\begin{flalign}
    \frac{dc_X}{dt} &= 0. \label{eq:stationary_condition}
\end{flalign}

\section{Results}
\label{sec:results}
\begin{figure*}[htbp]
    \centering
    \includegraphics[width=\linewidth]{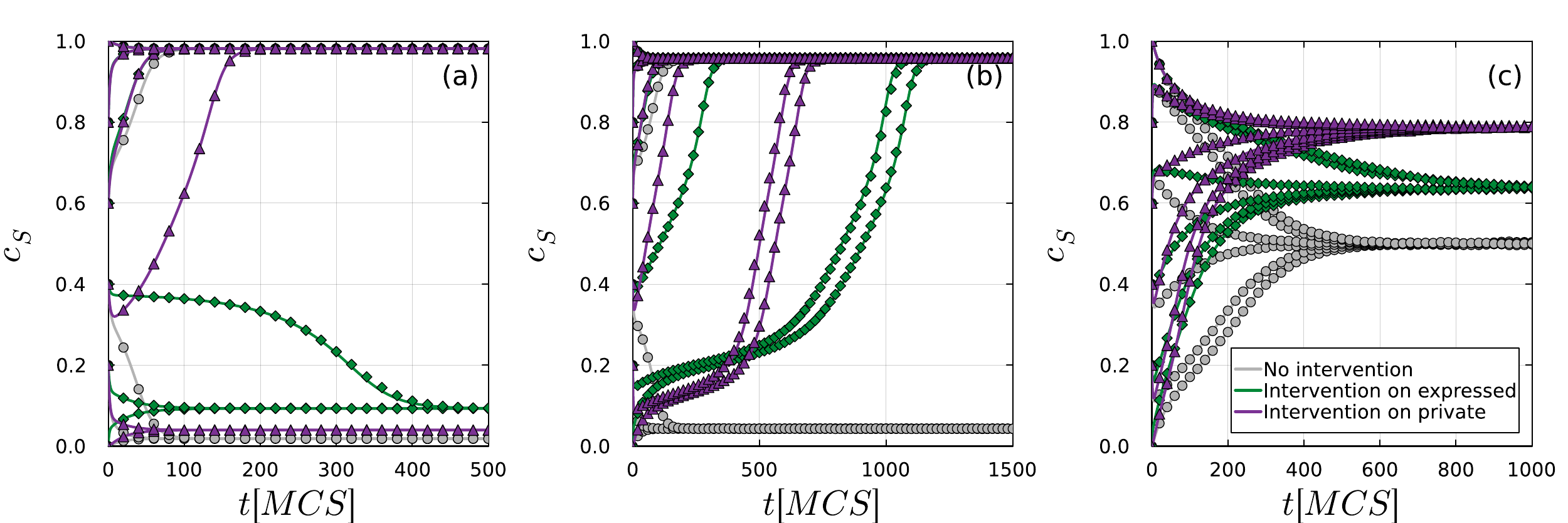}
    \caption{\textbf{Time evolution of the fraction of agents with a positive expressed opinion, $c_S$}, for three cases: the baseline without intervention ($h_e=h_p=0$), intervention applied to the expressed opinion ($h_e=0.05$, $h_p=0$), and intervention applied to the private opinion ($h_e=0$, $h_p=0.05$). Solid lines show the mean-field approximation, while markers show Monte Carlo simulations. Results are shown for influence group size $q=3$, probability of private-opinion update $\alpha=0.1$, and probability of independence: (a) $p=0.15$, (b) $p=0.2$, and (c) $p=0.28$.}
\label{fig:trajectories}    
\end{figure*}

We analyze the model under two distinct intervention strategies, applied to either the expressed or private level, and compare their outcomes to the baseline model without interventions, previously studied in \cite{kaminska_competition_2026}.
In all scenarios, after an initial transient period, the system reaches a stationary state in which the fraction of adopters, $c_S$, remains constant.
Based on this stationary value, we distinguish between two market-penetration regimes: 
a high-adoption regime, where the majority of agents are adopters ($c_S>0.5$), and a low-adoption regime, where  adopters form a minority ($c_S<0.5$).
The time evolution of $c_S$ towards these regimes is illustrated in Fig.~\ref{fig:trajectories}.

For low independence levels, Fig.~\ref{fig:trajectories} (a),
the system eventually reaches one of two possible stationary states corresponding to either the high- or low-adoption regime regardless of the intervention strategy. The initial conditions determine which regime is reached. 
In contrast to the baseline model,
where the stationary states are located symmetrically around $c_S=0.5$, the introduction of interventions breaks this symmetry, increasing the fraction of adopters in the low-adoption regime.
A bigger increase is achieved when the intervention acts at the expressed level.

For larger independence levels, Fig.~\ref{fig:trajectories} (b), 
the stationary state corresponding to the low-adoption regime disappears in models with interventions, leaving the high-adoption regime as the only stationary state. Thus, all trajectories converge to the high-adoption state, independently of the initial conditions. 
This convergence is faster when the intervention acts on the private level. However, both the interventions eventually lead to similar final fractions of adopters.
In contrast, in the baseline model, the low-adoption state can still be reached from certain initial conditions.

Finally, for even larger independence levels, Fig.~\ref{fig:trajectories} (c), 
applying the intervention to the private level leads to a noticeable higher stationary fraction of adopters than targeting the expressed level. 
Similar time trajectories can be obtained for the fraction of agents with positive private opinions, $c_\sigma$, and the level of cognitive dissonance, $d$.
However, since our primary interest lies in the long-run behavior of the system, we focus our analysis on the stationary states.

\begin{figure*}[htbp]
    \centering
    \includegraphics[width=\linewidth]{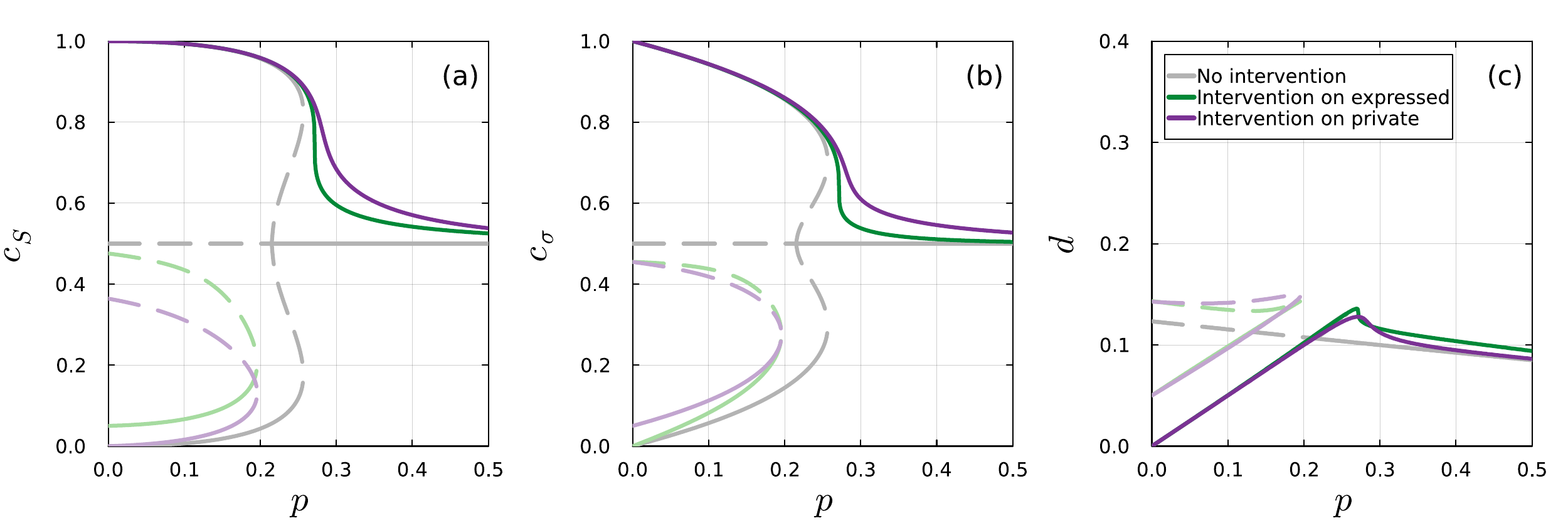}
    \caption{\textbf{Stationary concentrations of positive opinions at (a) expressed level $c_S$, (b) private level $c_\sigma$, and (c) dissonance $d$ as a function of the probability of independence $p$} obtained from mean field approximation with influence group of size $q = 3$, probability of private-opinion update $\alpha = 0.1$, probability of of intervention $h_e =0.0, h_p=0.0$, $h_e = 0.05, h_p = 0.0$ and $h_e =0.0, h_p = 0.05$. Solid lines indicate stable stationary states, that is, states toward which the system evolves from a range of initial conditions, whereas dashed lines indicate unstable stationary states. If we initialize the system exactly at such a state, it remains there; however, even a small change in the initial condition leads the system toward one of the stable stationary states.
    Vivid colored curves correspond to the high-adoption regime, whereas pale colored curves to the low-adoption regime.
    }
    \label{fig:phase_diagram_2025_alpha_0.1}
\end{figure*}

Figure~\ref{fig:phase_diagram_2025_alpha_0.1} illustrates how different interventions impact the stationary values of $c_S$, $c_\sigma$, and $d$. 
Our two market-penetration regimes are represented there by two distinct branches.
The high-adoption branch is depicted by vivid colors, whereas the low-adoption branch by pale colors.
Crucially, these two regimes are characterized by different levels of cognitive dissonance.
The agents in the high-adoption regime experience lower dissonance than those in the low-adoption branch, as shown in Fig.~\ref{fig:phase_diagram_2025_alpha_0.1} (c).
This happens regardless of the innervation type.
Such a result stands in contrast to the baseline model without interventions, where the dissonance remains the same across both the regimes.

For lower values of $p$, where two regimes are possible, the difference between the interventions is most visible in the low-adoption state. 
If the intervention acts on the expressed opinion, the stationary fraction of adopters, $c_S$, is higher than in the baseline case and also higher than in the case when the intervention acts on the private opinions, as seen in Fig.~\ref{fig:phase_diagram_2025_alpha_0.1} (a). In contrast, looking at the private level, Fig.~\ref{fig:phase_diagram_2025_alpha_0.1} (b) indicates that the stationary value of $c_\sigma$ is highest when the intervention acts directly on the private opinions. 
Thus, in the low-adoption state, the choice of the most effective intervention depends on whether the policymaker aims to increase public adoption or alter internal beliefs.
Acting on the expressed level is more effective at driving actual adoption, whereas acting on the private level successfully changes private beliefs but has a weaker effect on public adoption.

For higher values of $p$, the low-adoption state  disappears, and the differences between the interventions become visible in the high-adoption state. 
As $p$ increases, the stationary values of  $c_S$ and $c_\sigma$ decrease. However, this decrease is weaker when an intervention is present. 
In this case, the intervention acting on the private level maintains a higher fraction of adopters, see Figs.~\ref{fig:phase_diagram_2025_alpha_0.1} (a), as well as a higher fraction of positive private beliefs, see Figs.~\ref{fig:phase_diagram_2025_alpha_0.1} (b), than the intervention acting on the expressed level.
Thus, in the high-adoption regime, an intervention at the private level is more effective at promoting both public adoption and positive private beliefs.

Since varying the intervention intensities, $h_e$ and $h_p$, and the probability of private-opinion update, $\alpha$, do not change our qualitative conclusions, these additional results are presented in Appendix~\ref{app:interv_dependence}.

\section{Discussion}
\label{sec:discission}
In this section, we discuss the results in the context of consumer behavior. First, we relate the model outcomes to self-congruity theory, which has been used to explain consumer behavior in marketing research for the last five decades \citep{kolanskastronka_five_2024}. The agreement between the model results and this well-established theory provides a form of theoretical validation of the model. Second, we discuss the results in relation to the intention–behavior gap. This gap is directly connected with the main distinction used in the model, namely the distinction between private opinion and expressed behavior. It also helps to interpret why interventions that change internal acceptance and interventions that change visible behavior may have different effects.

\subsection{Dissonance and self-congruity theory}
In the model, the external intervention is defined abstractly as an influence that promotes adoption. An intervention acting on the expressed level can be interpreted as a tool that directly affects visible behavior or public declaration, for example advertising, a promotional campaign, a subsidy, a legal requirement, or another formal incentive. In contrast, an intervention acting on the private level can be interpreted as a tool that affects the internal evaluation of the product, practice, or behavior, for example advertising, education, information that changes beliefs, arguments that increase perceived usefulness, or communication that makes adoption more consistent with the consumer's self-image.

Across all model versions, the external intervention leads to the emergence of two market-penetration regimes, represented by two distinct branches in Fig.~\ref{fig:phase_diagram_2025_alpha_0.1}: a high-adoption branch with low dissonance and a low-adoption branch with higher dissonance. These two regimes can be related to different scenarios discussed in self-congruity theory \citep{sirgy_self-concept_1982,sirgy_revisiting_2016}, a well-established theory in marketing research \citep{kolanskastronka_five_2024}. Derived from psychology and behavioral sciences, this theory explains self-congruity as a cognitive evaluation process in which consumers compare the perceived image of a brand, product, or service with their own self-concept \citep{sirgy_self-concept_1982,sirgy_revisiting_2016}. This perspective fits our model, in which we distinguish between expressed adoption and private opinion and measure dissonance as the mismatch between the two.

In our model, the high-adoption branch is accompanied by low dissonance, which is consistent with self-congruity theory. In the language of self-congruity theory, this can be interpreted as a situation in which the promoted product, practice, or behavior fits how consumers see themselves or how they would like to see themselves \citep{marshall_endorsement_2008,sirgy_revisiting_2016}.
This consistency may reduce post-purchase dissonance and support repeat-purchase behavior, which helps sustain higher adoption levels.
On the other hand, the low-adoption branch is accompanied by higher dissonance and can be related to a mismatch between the promoted image and consumers' self-concept. In such a case, cognitive dissonance may turn into regret \citep{shahin_sharifi_impacts_2014}. This decreases decision satisfaction and increases brand-switching intention, consequently preventing the system from reaching a high level of market penetration \citep{bui_modeling_2011}. Hence, the results of our model are consistent with a well-established theory of consumer behavior.

The above interpretation suggests that external intervention should not only encourage adoption, but also reduce the risk of dissonance after adoption. Previous studies on post-purchase dissonance suggest that this can be supported by post-purchase communication that helps consumers justify their choice \citep{shahin_sharifi_impacts_2014}. Such communication may include additional product information, care instructions, or performance reassurance, which can function as consonant cognitions \citep{mcgrath_dealing_2017}. Furthermore, advertising should remain congruent with the consumer's actual or ideal self-concept \citep{marshall_endorsement_2008,sirgy_revisiting_2016}. If the projected product image is too far from the consumer image, it can increase the dissonance. In the context of our model, such an incongruent image acts as a repulsive force, potentially pushing the system out of the high-adoption regime into the lower branch of market penetration.

\subsection{Intention-Behavior Gap}
The attitude-behavior gap can explain why some kind of interventions are more effective than others. For example, in a filed study on household energy conservation, \cite{asensio_nonprice_2015} found that the participants when surveyed consistently rank savings as their top concern.
However, these monetary incentives failed to drive the actual change, leaving the system in the low-adoption branch due to high dissonance.
In fact, monetary incentives may be counterproductive by undermining more altruistic or prosocial motivations \cite{delmas_information_2013}.
In contrast, environmental and health-based education campaigns resulted in significant energy conservation by directly targeting the consumer ideal self-concept, which turned out to be particularly effective on families with children \cite{asensio_nonprice_2015}.
By framing energy conservation as a moral duty to reduce health risks, the education campaign acts as a powerful consonant cognition. 
This reduces cognitive dissonance, allowing consumers to align their behavior with their moral self-image, which ultimately stabilizes the system in the high-adoption branch.
Similarly, educational interventions were more effective than monetary incentives in reducing water consumption \cite{rajapaksa_monetary_2019}.

Although the above examples demonstrate a clear advantage of educational campaigns, field studies on promoting heathy food purchases are less conclusive.
The nutrition education alone has yielded mixed results, showing positive impact in some studies \cite{anderson_5_2001,bihan_impact_2011} but zero impact in others \cite{an_effectiveness_2013}. 
This may suggest that education champagnes only succeed when they act successfully as consonant cognitions. If the message is poorly designed or too far from the consumer ideal self-image, they fail to reduce dissonance.
In contrast, food subsidies increase healthy food consumption by lowering financial barriers \cite{an_effectiveness_2013}.
This aligns with previous findings indicating that reducing adoption costs and barriers can promote the diffusion of new products and practices \cite{byrka_difficulty_2016}.
Finally, some studies show that the combination of both monetary and non-monetary strategies yields the best effects \cite{anderson_5_2001,streletskaya_taxes_2013}.

\section{Conclusions}
\label{sec:conclusions}
In this paper, we have extended the expressed-private opinion model by introducing an external intervention that can act either on the expressed or on the private level. This allowed us to address the question posed in the Introduction: should an intervention target visible behavior or internal acceptance (intention)? The answer is: it depends. More precisely, it depends on the adoption regime in which the system evolves. If adoption is low, intervention at the expressed level is more effective in increasing the fraction of visible adopters. Such interventions may be interpreted as subsidies, regulations, promotional campaigns, or other incentives that directly affect behavior. If adoption is already high, intervention at the private level is more effective because it supports internal acceptance and reduces dissonance. Such interventions may be interpreted as education, persuasive information, or communication that makes adoption more consistent with consumers’ values and self-image.

This result leads to a simple policy implication. At early stages of diffusion, expressed-level interventions may help to initiate adoption and make the behavior more visible. At later stages, private-level interventions may be needed to stabilize adoption and reduce the gap between what agents do publicly and what they accept privately. Therefore, the two types of intervention should not necessarily be treated as alternatives. Rather, they may play different roles at different stages of the diffusion process.

The emergent outcome of our model, which increases its credibility, is the appearance of two adoption branches: a high-adoption branch with low dissonance and a low-adoption branch with higher dissonance. These branches were not imposed by the model assumptions, but arise from the dynamics, and their interpretation is consistent with self-congruity theory. In particular, high adoption accompanied by low dissonance agrees with the idea that adoption is more stable when public behavior and private acceptance are aligned. Thus, the model is not only internally consistent, as confirmed by the agreement between mean-field approximation and Monte Carlo simulations, but also qualitatively consistent with an established psychological theory.

In spite of this agreement with self-congruity theory, we do not claim that the present model is the ultimate description of intervention-driven adoption. We are aware that we have answered the main question only within one family of models. The proposed model is very simple, which is a good practice as the first step in agent-based modeling. Such simplicity allows us to isolate the mechanism and understand its consequences. However, to check the robustness of our results, it would be desirable to extend the analysis to broader classes of models, including other update rules, heterogeneous agents, more realistic network structures.

\section*{Acknowledgments}
Funded by the National Science Centre, Poland under the OPUS call in the Weave programme, project no. 2023/51/I/HS6/02269.

\section*{Declaration of generative AI and AI-assisted technologies in the manuscript preparation process}
During the preparation of this work, the authors used ChatGPT, developed by OpenAI, to assist with language correction and wording suggestions. The authors reviewed and edited the output as needed and take full responsibility for the content of the published article.

\appendix

\section{Elementary update algorithm}
\label{app:algorithm}
The elementary update is performed according to the following algorithm:

\begin{algorithm}[H]
\caption{Elementary update}
\label{alg:opinion_update}
\begin{algorithmic}[1]
\State Select a target agent $i$ uniformly at random from the population
\State Draw a random value $r_{\alpha} \sim U(0,1)$

\If{$r_{\alpha} < \alpha$} \Comment{\textbf{Private Opinion Level Update}}
    \State Draw a random value $r_p \sim U(0,1)$
    \If{$r_p < p$} \Comment{Independence}
        \State Draw a random value $r_{indep}\sim U(0,1)$
        \If{$r_{indep} < 0.5$}
            \State $\sigma_i \gets -\sigma_i$
        \EndIf
    \Else \Comment{Susceptible to Influence}
        \State Draw a random value $r_h \sim U(0,1)$
        \If{$r_h < h_p$} \Comment{External Intervention}
            \State $\sigma_i \gets 1$
        \Else \Comment{Social Influence ($q$-panel)}
            \State Select $q$ agents $\{i_1, i_2, \dots, i_q\}$ uniformly at random from the neighbors of the target agent
            \If{$S_i = S_{i_1} = S_{i_2} = \dots = S_{i_q}$}
                \State $\sigma_i \gets S_i$
            \EndIf
        \EndIf
    \EndIf

\Else \Comment{\textbf{Expressed Opinion Level Update}}
    \State Draw a random value $r_p \sim U(0,1)$
    \If{$r_p < p$} \Comment{Independence}
        \State $S_i \gets \sigma_i$
    \Else \Comment{Susceptible to Influence}
        \State Draw a random value $r_h \sim U(0,1)$
        \If{$r_h < h_e$} \Comment{External Intervention}
            \State $S_i \gets 1$
        \Else \Comment{Social Influence ($q$-panel)}
            \State Select $q$ agents $\{i_1, i_2, \dots, i_q\}$ uniformly at random from the neighbors of the target agent
            \If{$S_i \neq \sigma_i$ \textbf{and} $\exists j \in \{1, \dots, q\}$ such that $S_{i_j} = \sigma_i$}
                \State $S_i \gets \sigma_i$
            \ElsIf{$S_i = \sigma_i$ \textbf{and} $S_{i_1} = S_{i_2} = \dots = S_{i_q}$}
                \State $S_i \gets S_{i_1}$
            \EndIf
        \EndIf
    \EndIf
\EndIf
\end{algorithmic}
\end{algorithm}

\section{Mean-field approximation equations}
\label{app:MFA}
As stated in Sec.~\ref{sec:methods}, the time evolution of the model is described by the system of nonlinear differential equations \eqref{eq:rate}.  
These equations have the following forms:
\begin{flalign}
    \frac{dc_{\uparrow \uparrow}}{dt} &=c_{\sigma} (1-\alpha) [1 - (1-p)(1 - h_e)(1-c_S)^q] \nonumber \\
    & + c_{\uparrow \uparrow}\alpha \left[1-p/2 \, \right] \nonumber \\
    & + c_{\uparrow\downarrow}\alpha \left[p/2+(1-p)[ h_p + (1-h_p) c_S^q] \right] \nonumber \\
    & - c_{\uparrow \uparrow} ,
    \label{eq:c_up_up_2025} \\
    \frac{dc_{\uparrow \downarrow}}{dt} &= (1-c_{\sigma})(1-\alpha)(1-p)\left[h_e + (1-h_e) c_S^q \right] \nonumber \\
    &  + c_{\uparrow \uparrow}\alpha\left[p/2 \, \right]  \nonumber \\
    &+ c_{\uparrow\downarrow}\alpha\left[1 - p/2 - (1-p)[ h_p + (1-h_p)c_S^q]\right] \nonumber \\
    & - c_{\uparrow \downarrow},
    \label{eq:c_up_down_2025} \\    
    \frac{dc_{\downarrow \uparrow}}{dt} &= c_{\sigma}(1-\alpha)(1-p)(1-h_e)(1-c_S)^q  \nonumber \\
    & + c_{\downarrow\uparrow}\alpha[1-p/2-(1-p)(1-h_p)(1-c_S)^q]   \nonumber \\
    &+ c_{\downarrow\downarrow}\alpha\left[p/2 + (1-p)h_p \, \right]\nonumber \\
    & - c_{\downarrow \uparrow},  
    \label{eq:c_down_up_2025} \\ 
    \frac{dc_{\downarrow \downarrow}}{dt} &= (1-c_{\sigma})(1-\alpha)\left[1-(1-p)[h_e + (1-h_e) c_S^q]\right]  \nonumber \\
    & + c_{\downarrow\uparrow}\alpha[p/2+(1-p)(1-h_p)(1-c_S)^q]   \nonumber \\
    &+ c_{\downarrow\downarrow}\alpha\left[1 - p/2 -(1-p)h_p  \, \right] \nonumber \\
    & - c_{\downarrow \downarrow}.
    \label{eq:c_down_down_2025}
\end{flalign}

We can only solve these equations numerically and such solutions are presented in Secs.~\ref{sec:results} and \ref{app:interv_dependence}. All MATLAB codes used to numerically solve these systems of equations are available on Zenodo \citep{kaminska_2026_zenodo}. 

\section{Dependence on intervention intensity $h_{e,p}$ and private-opinion update $\alpha$}
\label{app:interv_dependence}

\begin{figure*}[htbp]
    \centering
    \includegraphics[width=\linewidth]{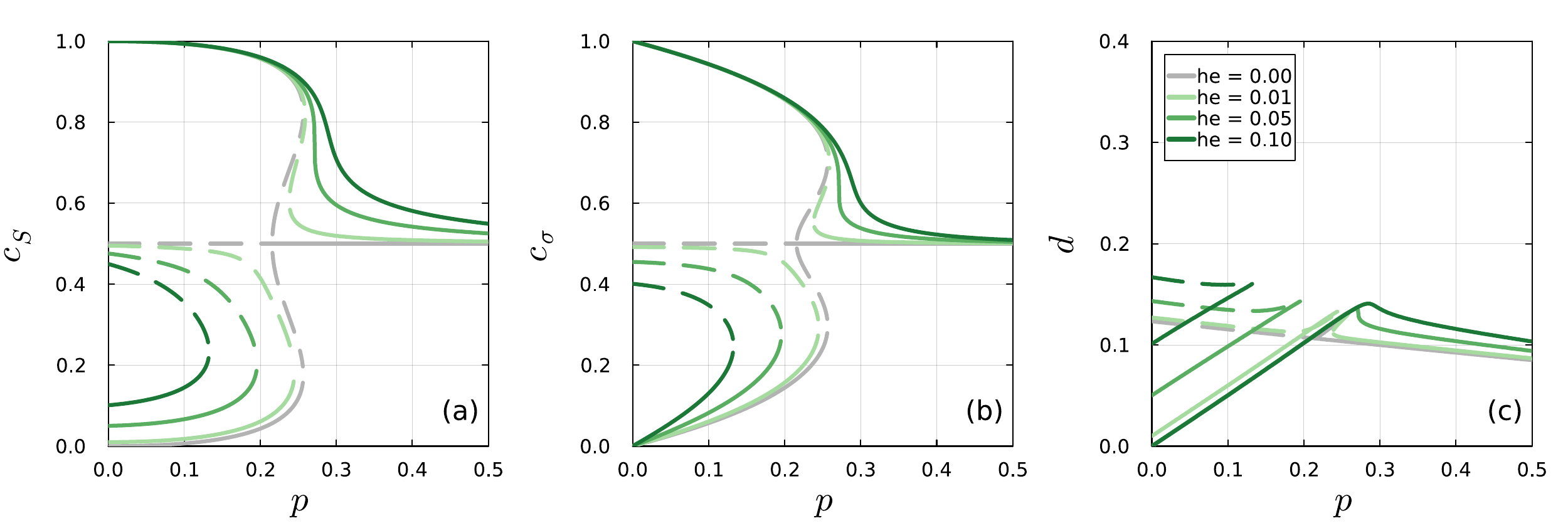}
    \includegraphics[width=\linewidth]{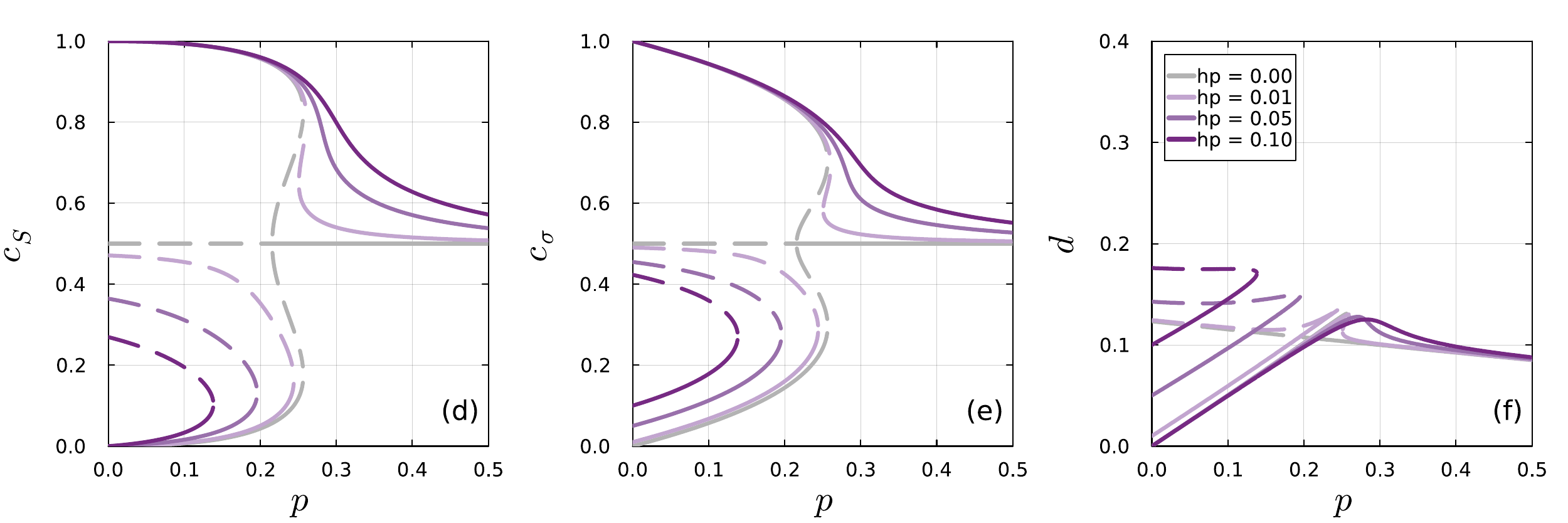}
    \caption{Stationary concentrations of positive opinions at expressed level $c_S$ (left), private level $c_\sigma$ (middle) and dissonance $d$ (right) as a function of the probability of independence $p$ obtained from mean field approximation with influence group of size $q = 3$, probability of private-opinion update $\alpha = 0.1$,. Darker color indicates increasing probability of intervention $h_e = 0.0, 0.01, 0.05, 0.1, h_p = 0$ (upper) and $h_e = 0, h_p = 0.0, 0.01, 0.05, 0.1$ (lower). }
    \label{fig:h_dep_2025}
\end{figure*}

\begin{figure*}[htbp]
    \centering
    \includegraphics[width=\linewidth]{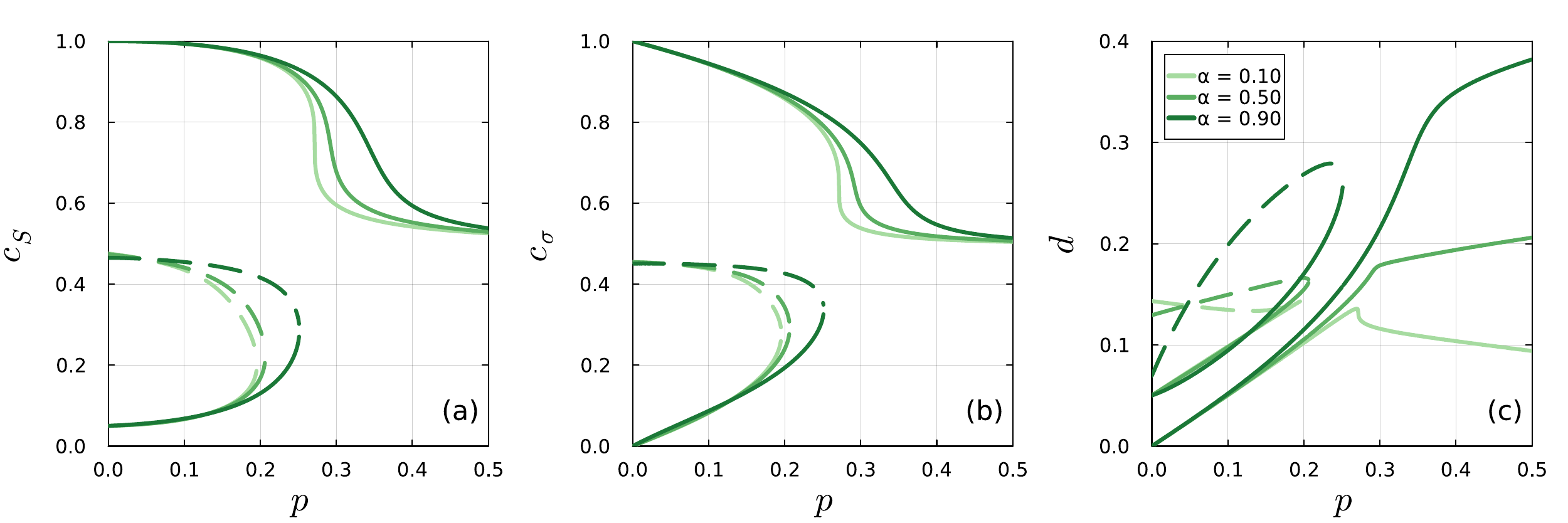}
    \includegraphics[width=\linewidth]{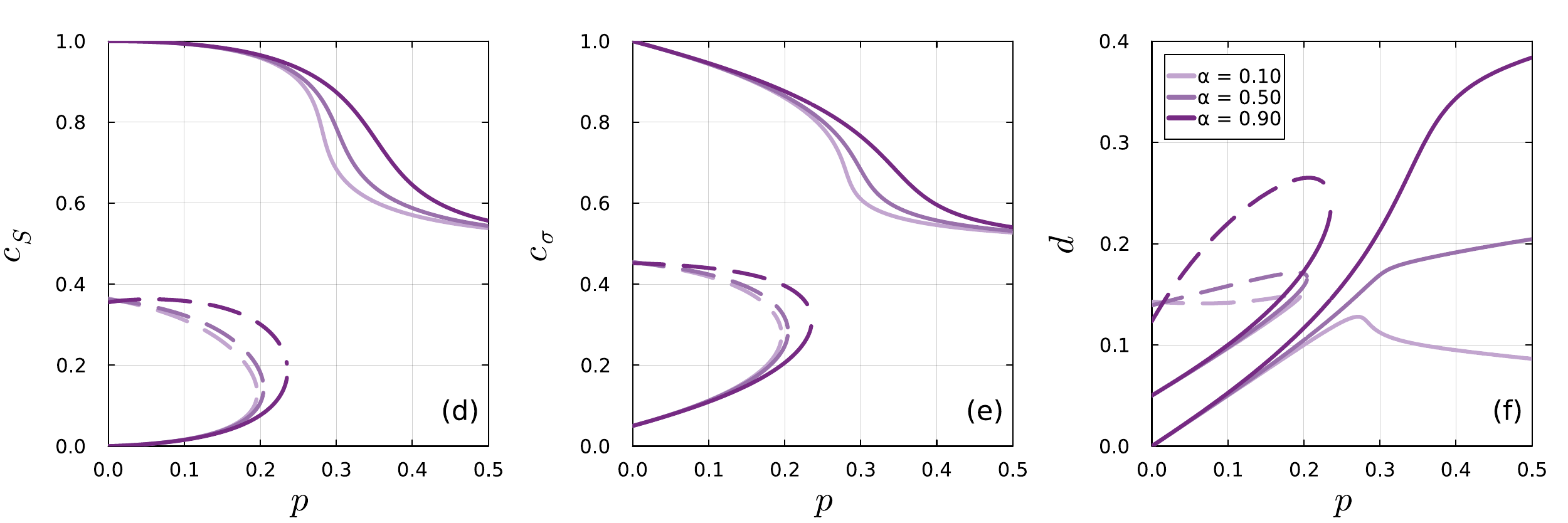}
    \caption{Stationary concentrations of positive opinions at expressed level $c_S$ (left), private level $c_\sigma$ (middle) and dissonance $d$ (right) as a function of the probability of independence $p$ obtained from mean field approximation with influence group of size $q = 3$, probability of intervention $h_e =0.05, h_p = 0.0$ (upper) and $h_e = 0.0, h_p = 0.05$ (lower). Darker color indicates increasing probability of private-opinion update $\alpha = 0.1, 0.5, 0.9$.}
    \label{fig:alpha_dep_2025}
\end{figure*}

\bibliographystyle{elsarticle-harv}
\bibliography{references}

@article{kaminska_competition_2026,
	title = {Competition between private and expressed opinions in binary choice: The $\alpha$-{EPO} $q$-voter model},
	volume = {113},
	issn = {2470-0045, 2470-0053},
	shorttitle = {Competition between private and expressed opinions in binary choice},
	doi = {10.1103/8w9t-jpvp},
	language = {en},
	number = {6},
	journal = {Physical Review E},
	author = {Kamińska, Barbara and Nowak, Barbara and Lipiecki, Arkadiusz and Sznajd-Weron, Katarzyna},
	month = jun,
	year = {2026},
	pages = {064301},
}

@article{dong_opinion_2024,
	title = {Opinion formation analysis for {Expressed} and {Private} {Opinions} ({EPOs}) models: {Reasoning} private opinions from behaviors in group decision-making systems},
	volume = {236},
	issn = {09574174},
	shorttitle = {Opinion formation analysis for {Expressed} and {Private} {Opinions} ({EPOs}) models},
	doi = {10.1016/j.eswa.2023.121292},
	language = {en},
	journal = {Expert Systems with Applications},
	author = {Dong, Jianglin and Hu, Jiangping and Zhao, Yiyi and Peng, Yuan},
	month = feb,
	year = {2024},
	pages = {121292},
}

@article{convery_most_2007,
	title = {The most popular tax in {Europe}? {Lessons} from the {Irish} plastic bags levy},
	volume = {38},
	copyright = {http://www.springer.com/tdm},
	issn = {0924-6460, 1573-1502},
	shorttitle = {The most popular tax in {Europe}?},
	doi = {10.1007/s10640-006-9059-2},
	language = {en},
	number = {1},
	journal = {Environmental and Resource Economics},
	author = {Convery, Frank and McDonnell, Simon and Ferreira, Susana},
	month = jul,
	year = {2007},
	pages = {1--11},
}

@article{truelove_curbing_2022,
	title = {Curbing single-use plastic with behaviour change interventions},
	volume = {3},
	copyright = {2022 Springer Nature Limited},
	issn = {2662-138X},
	doi = {10.1038/s43017-022-00356-y},
	language = {en},
	number = {11},
	journal = {Nature Reviews Earth \& Environment},
	author = {Truelove, Heather Barnes and Raimi, Kaitlin T. and Carrico, Amanda R.},
	month = nov,
	year = {2022},
	pages = {722--723},
}

@article{sirgy_self-concept_1982,
	title = {Self-{Concept} in {Consumer} {Behavior}: {A} {Critical} {Review}},
	volume = {9},
	issn = {0093-5301, 1537-5277},
	shorttitle = {Self-{Concept} in {Consumer} {Behavior}},
	doi = {10.1086/208924},
	language = {en},
	number = {3},
	journal = {Journal of Consumer Research},
	author = {Sirgy, M. Joseph},
	month = dec,
	year = {1982},
	pages = {287},
}

@article{kolanskastronka_five_2024,
	title = {Five decades of self‐congruity in consumer behaviour research: {A} systematic review and future research agenda},
	volume = {48},
	issn = {1470-6423, 1470-6431},
	shorttitle = {Five decades of self‐congruity in consumer behaviour research},
	doi = {10.1111/ijcs.12983},
	language = {en},
	number = {1},
	journal = {International Journal of Consumer Studies},
	author = {Kolańska‐Stronka, Magdalena and Singh, Balgopal},
	month = jan,
	year = {2024},
	pages = {e12983},
}

@article{byrka_difficulty_2016,
	title = {Difficulty is critical: {The} importance of social factors in modeling diffusion of green products and practices},
	volume = {62},
	issn = {1364-0321},
	shorttitle = {Difficulty is critical},
	doi = {10.1016/j.rser.2016.04.063},
	journal = {Renewable and Sustainable Energy Reviews},
	author = {Byrka, Katarzyna and J\k{e}drzejewski, Arkadiusz and Sznajd-Weron, Katarzyna and Weron, Rafał},
	month = sep,
	year = {2016},
	pages = {723--735},
}

@article{rajapaksa_monetary_2019,
	title = {Do monetary and non-monetary incentives influence environmental attitudes and behavior? {Evidence} from an experimental analysis},
	volume = {149},
	issn = {0921-3449},
	shorttitle = {Do monetary and non-monetary incentives influence environmental attitudes and behavior?},
	doi = {10.1016/j.resconrec.2019.05.034},
	journal = {Resources, Conservation and Recycling},
	author = {Rajapaksa, Darshana and Gifford, Robert and Torgler, Benno and Garcia-Valiñas, Marian and Athukorala, Wasantha and Managi, Shunsuke and Wilson, Clevo},
	month = oct,
	year = {2019},
	pages = {168--176},
}

@article{castellano_statistical_2009,
	title = {Statistical physics of social dynamics},
	volume = {81},
	copyright = {http://link.aps.org/licenses/aps-default-license},
	issn = {0034-6861, 1539-0756},
	doi = {10.1103/RevModPhys.81.591},
	language = {en},
	number = {2},
	journal = {Reviews of Modern Physics},
	author = {Castellano, Claudio and Fortunato, Santo and Loreto, Vittorio},
	month = may,
	year = {2009},
	pages = {591--646},
}

@article{nail_proposal_2000,
	title = {Proposal of a four-dimensional model of social response.},
	volume = {126},
	issn = {1939-1455, 0033-2909},
	doi = {10.1037/0033-2909.126.3.454},
	language = {en},
	number = {3},
	journal = {Psychological Bulletin},
	author = {Nail, Paul R. and MacDonald, Geoff and Levy, David A.},
	year = {2000},
	pages = {454--470},
}

@article{rausch_bridge_2021,
	title = {Bridge the gap: {Consumers}’ purchase intention and behavior regarding sustainable clothing},
	volume = {278},
	issn = {09596526},
	shorttitle = {Bridge the gap},
	doi = {10.1016/j.jclepro.2020.123882},
	language = {en},
	journal = {Journal of Cleaner Production},
	author = {Rausch, Theresa Maria and Kopplin, Cristopher Siegfried},
	month = jan,
	year = {2021},
	pages = {123882},
}

@article{park_exploring_2020,
	title = {Exploring attitude–behavior gap in sustainable consumption: comparison of recycled and upcycled fashion products},
	volume = {117},
	issn = {01482963},
	shorttitle = {Exploring attitude–behavior gap in sustainable consumption},
	doi = {10.1016/j.jbusres.2018.08.025},
	language = {en},
	journal = {Journal of Business Research},
	author = {Park, Hyun Jung and Lin, Li Min},
	month = sep,
	year = {2020},
	pages = {623--628},
}

@article{kaminska_impact_2025,
	title = {Impact of cognitive dissonance on social hysteresis: {Insights} from the expressed and private opinions model},
	volume = {273},
	issn = {09574174},
	shorttitle = {Impact of cognitive dissonance on social hysteresis},
	doi = {10.1016/j.eswa.2025.126851},
	language = {en},
	journal = {Expert Systems with Applications},
	author = {Kamińska, Barbara and Sznajd-Weron, Katarzyna},
	month = may,
	year = {2025},
	pages = {126851},
}

@article{jedrzejewski_think_2018,
	title = {Think then act or act then think?},
	volume = {13},
	issn = {1932-6203},
	doi = {10.1371/journal.pone.0206166},
	language = {en},
	number = {11},
	journal = {PLOS ONE},
	author = {J\k{e}drzejewski, Arkadiusz and Marcjasz, Grzegorz and Nail, Paul R. and Sznajd-Weron, Katarzyna},
	editor = {Gonçalves, Sebastián},
	month = nov,
	year = {2018},
	pages = {e0206166},
}

@article{nieto-garcia_consumer_2024,
	title = {Consumer hypocrisy and researcher myopia: {A} scrutiny of the intention-behaviour gap in sustainable tourism},
	volume = {104},
	issn = {01607383},
	shorttitle = {Consumer hypocrisy and researcher myopia},
	doi = {10.1016/j.annals.2023.103678},
	language = {en},
	journal = {Annals of Tourism Research},
	author = {Nieto-García, Marta and Acuti, Diletta and Viglia, Giampaolo},
	month = jan,
	year = {2024},
	pages = {103678},
}

@article{de_sio_product_2024,
	title = {Product {Characteristics} and {Emotions} to {Bridge} the {Intention}-{Behavior} {Gap} in {Green} {Food} {Purchasing}},
	volume = {16},
	issn = {2071-1050},
	doi = {10.3390/su16177297},
	language = {en},
	number = {17},
	journal = {Sustainability},
	author = {De Sio, Sara and Casu, Giulia and Zamagni, Alessandra and Gremigni, Paola},
	month = aug,
	year = {2024},
	pages = {7297},
}

@article{linder_psychological_2026,
	title = {Psychological barriers for sustainable diets: {Unpacking} intention-behavior gaps in meat consumption},
	volume = {135},
	issn = {09503293},
	shorttitle = {Psychological barriers for sustainable diets},
	doi = {10.1016/j.foodqual.2025.105721},
	language = {en},
	journal = {Food Quality and Preference},
	author = {Linder, Noah and Lindahl, Therse and Wijermans, Nanda},
	month = jan,
	year = {2026},
	pages = {105721},
}

@article{an_effectiveness_2013,
	title = {Effectiveness of subsidies in promoting healthy food purchases and consumption: a review of field experiments},
	volume = {16},
	issn = {1368-9800, 1475-2727},
	shorttitle = {Effectiveness of subsidies in promoting healthy food purchases and consumption},
	doi = {10.1017/S1368980012004715},
	language = {en},
	number = {7},
	journal = {Public Health Nutrition},
	author = {An, Ruopeng},
	month = jul,
	year = {2013},
	pages = {1215--1228},
}

@article{anderson_5_2001,
  title = {5 A Day Fruit and Vegetable Intervention Improves Consumption in a Low Income Population},
  volume = {101},
  ISSN = {0002-8223},
  DOI = {10.1016/s0002-8223(01)00052-9},
  number = {2},
  journal = {Journal of the American Dietetic Association},
  publisher = {Elsevier BV},
  author = {Anderson, Judith V. and Bybee, Deborah I. and Brown, Randi M. and McLean, Donna F. and Garcia, Erika M. and Breer, M. Lynn and Schillo, Barbara A.},
  year = {2001},
  month = Feb,
  pages = {195–202}
}

@article{bihan_impact_2011,
  title = {Impact of fruit and vegetable vouchers and dietary advice on fruit and vegetable intake in a low-income population},
  volume = {66},
  ISSN = {1476-5640},
  DOI = {10.1038/ejcn.2011.173},
  number = {3},
  journal = {European Journal of Clinical Nutrition},
  publisher = {Springer Science and Business Media LLC},
  author = {Bihan,  H and Méjean,  C and Castetbon,  K and Faure,  H and Ducros,  V and Sedeaud,  A and Galan,  P and Le Clésiau,  H and Péneau,  S and Hercberg,  S},
  year = {2011},
  month = Oct,
  pages = {369–375}
}

@article{streletskaya_taxes_2013,
  title = {Taxes,  Subsidies,  and Advertising Efficacy in Changing Eating Behavior: An Experimental Study},
  volume = {36},
  ISSN = {2040-5804},
  DOI = {10.1093/aepp/ppt032},
  number = {1},
  journal = {Applied Economic Perspectives and Policy},
  publisher = {Wiley},
  author = {Streletskaya,  Nadia A. and Rusmevichientong,  Pimbucha and Amatyakul,  Wansopin and Kaiser,  Harry M.},
  year = {2013},
  month = Nov,
  pages = {146–174}
}

@article{asensio_nonprice_2015,
	title = {Nonprice incentives and energy conservation},
	volume = {112},
	doi = {10.1073/pnas.1401880112},
	number = {6},
	journal = {Proceedings of the National Academy of Sciences},
	publisher = {Proceedings of the National Academy of Sciences},
	author = {Asensio, Omar I. and Delmas, Magali A.},
	month = feb,
	year = {2015},
	pages = {E510--E515},
}

@article{delmas_information_2013,
  title = {Information strategies and energy conservation behavior: A meta-analysis of experimental studies from 1975 to 2012},
  volume = {61},
  ISSN = {0301-4215},
  DOI = {10.1016/j.enpol.2013.05.109},
  journal = {Energy Policy},
  publisher = {Elsevier BV},
  author = {Delmas,  Magali A. and Fischlein,  Miriam and Asensio,  Omar I.},
  year = {2013},
  month = Oct,
  pages = {729–739}
}

@article{elhaffar_towards_2020,
	title = {Towards closing the attitude-intention-behavior gap in green consumption: {A} narrative review of the literature and an overview of future research directions},
	volume = {275},
	issn = {09596526},
	shorttitle = {Towards closing the attitude-intention-behavior gap in green consumption},
	doi = {10.1016/j.jclepro.2020.122556},
	language = {en},
	journal = {Journal of Cleaner Production},
	author = {ElHaffar, Ghina and Durif, Fabien and Dubé, Laurette},
	month = dec,
	year = {2020},
	pages = {122556},
}

@incollection{sirgy_revisiting_2016,
	title = {Revisiting self-congruity theory in consumer behaviour: Making sense of the research so far},
	booktitle = {Routledge international handbook of consumer psychology},
	publisher = {Routledge},
	author = {Sirgy, M. Joseph and Lee, Dong-Jin and Yu, Grace B.},
	year = {2016},
    doi = {10.4324/9781315727448-23},
    isbn = {9781138915886},
    pages = {185--201},
}

@article{shahin_sharifi_impacts_2014,
	title = {The impacts of relationship marketing on cognitive dissonance, satisfaction, and loyalty: {The} mediating role of trust and cognitive dissonance},
	volume = {42},
	copyright = {https://www.emerald.com/insight/site-policies},
	issn = {0959-0552},
	shorttitle = {The impacts of relationship marketing on cognitive dissonance, satisfaction, and loyalty},
	doi = {10.1108/IJRDM-05-2013-0109},
	language = {en},
	number = {6},
	journal = {International Journal of Retail \& Distribution Management},
	author = {Shahin Sharifi, Seyed and Rahim Esfidani, Mohammad},
	month = jun,
	year = {2014},
	pages = {553--575},
}

@article{mcgrath_dealing_2017,
	title = {Dealing with dissonance: {A} review of cognitive dissonance reduction},
	volume = {11},
	issn = {1751-9004, 1751-9004},
	shorttitle = {Dealing with dissonance},
	doi = {10.1111/spc3.12362},
	language = {en},
	number = {12},
	journal = {Social and Personality Psychology Compass},
	author = {McGrath, April},
	month = dec,
	year = {2017},
	pages = {e12362},
}

@article{marshall_endorsement_2008,
	title = {Endorsement {Theory}: {How} {Consumers} {Relate} to {Celebrity} {Models}},
	volume = {48},
	issn = {0021-8499, 1740-1909},
	shorttitle = {Endorsement {Theory}},
	doi = {10.2501/S0021849908080550},
	language = {en},
	number = {4},
	journal = {Journal of Advertising Research},
	author = {Marshall, Roger and Na, Woonbong and State, Gabriel and Deuskar, Sonali},
	month = dec,
	year = {2008},
	pages = {564--572},
}

@article{bui_modeling_2011,
	title = {Modeling regret effects on consumer post‐purchase decisions},
	volume = {45},
	copyright = {https://www.emerald.com/insight/site-policies},
	issn = {0309-0566},
	doi = {10.1108/03090561111137615},
	language = {en},
	number = {7/8},
	journal = {European Journal of Marketing},
	author = {Bui, My and Krishen, Anjala S. and Bates, Kenneth},
	month = jul,
	year = {2011},
	pages = {1068--1090},
}

@article{abrahamse_social_2013,
	title = {Social influence approaches to encourage resource conservation: {A} meta-analysis},
	volume = {23},
	issn = {09593780},
	shorttitle = {Social influence approaches to encourage resource conservation},
	doi = {10.1016/j.gloenvcha.2013.07.029},
	language = {en},
	number = {6},
	journal = {Global Environmental Change},
	author = {Abrahamse, Wokje and Steg, Linda},
	month = dec,
	year = {2013},
	pages = {1773--1785},
}

@misc{kaminska_2026_zenodo,
  author       = {Kamińska, Barbara},
  title        = {Expressed-Private opinions $q$-voter model with external interventions},
  month        = jun,
  year         = 2026,
  publisher    = {Zenodo},
  doi          = {10.5281/zenodo.20507468},
}



\end{document}